\documentclass[reprint, superscriptaddress, secnumarabic, amssymb, nobibnotes, aps, prl]{revtex4-1}

\usepackage{graphicx}
\usepackage{epstopdf}
\usepackage[T1]{fontenc}
\usepackage[latin9]{inputenc}
\usepackage{amsbsy}
\usepackage{gensymb}
\usepackage[T1]{fontenc}
\usepackage[latin9]{inputenc}
\usepackage{amsmath}
\usepackage{amssymb}
\usepackage{bbm}
\usepackage{braket}
\usepackage{xcolor}
\allowdisplaybreaks
\usepackage{graphicx}
\usepackage[colorlinks=true]{hyperref}  
\hypersetup{
    bookmarks=true,         
    unicode=false,          
    pdftoolbar=true,        
    pdfmenubar=true,        
    pdffitwindow=false,     
    pdfstartview={FitH},    
    pdftitle={An investigation of type I superconductivity in single crystal of Pb$_{2}$Pd },    
    pdfauthor={Arushi, R. P. Singh},     
    pdfsubject={},   
    pdfcreator={},   
    pdfproducer={}, 
    pdfkeywords={} {} {}, 
    pdfnewwindow=true,      
    colorlinks=true,       
    linkcolor=blue, 
    citecolor=blue,        
    filecolor=magenta,      
    urlcolor=blue           
} 
\usepackage[normalem]{ulem}

\newcommand{\equref}[1]{Eq.~(\ref{#1})}

\newcommand{\figref}[1]{Fig.~\ref{#1}}

\begin{document}
\title{\textrm{An investigation of type-I superconductivity in single crystal of Pb$_{2}$Pd }}
\author{Arushi}
\affiliation{Department of Physics, Indian Institute of Science Education and Research Bhopal, Bhopal, 462066, India}
\author{K.~Motla}
\affiliation{Department of Physics, Indian Institute of Science Education and Research Bhopal, Bhopal, 462066, India}
\author{A. Kataria}
\affiliation{Department of Physics, Indian Institute of Science Education and Research Bhopal, Bhopal, 462066, India}
\author{S. Sharma}
\affiliation{Department of Physics and Astronomy, McMaster University, Hamilton, Ontario L8S 4M1, Canada}
\author{J.~Beare}
\affiliation{Department of Physics and Astronomy, McMaster University, Hamilton, Ontario L8S 4M1, Canada}
\author{M.~Pula}
\affiliation{Department of Physics and Astronomy, McMaster University, Hamilton, Ontario L8S 4M1, Canada}  
\author{M.~Nugent}
\affiliation{Department of Physics and Astronomy, McMaster University, Hamilton, Ontario L8S 4M1, Canada}
\author{G.~M.~Luke}
\affiliation{Department of Physics and Astronomy, McMaster University, Hamilton, Ontario L8S 4M1, Canada}
\affiliation{TRIUMF, Vancouver, British Columbia V6T 2A3, Canada}
\author{R. P. Singh}
\email[]{rpsingh@iiserb.ac.in}
\affiliation{Department of Physics, Indian Institute of Science Education and Research Bhopal, Bhopal, 462066, India}
\date{\today}
\begin{abstract}
\begin{flushleft}
\end{flushleft}
We have investigated the superconducting properties in a single crystal of a new superconductor Pb$_{2}$Pd via various techniques including magnetization, AC transport, transverse field muon spin rotation and relaxation (TF-$\mu$SR), and heat capacity. Pb$_{2}$Pd crystallizes in a body-centred tetragonal structure with space group I4/$mcm$. All measurements confirm the superconducting transition temperature, T$_{C}$ = 3.0 $\pm$ 0.1 K. Electronic specific heat data are well described by the BCS fitting, suggesting that Pb$_{2}$Pd opens an isotropic gap on entering the superconducting state. The specific heat jump and $\lambda_{e-ph}$ value categorize Pb$_{2}$Pd as a moderately coupled superconductor. Magnetization and transverse field muon spin rotation measurements along with Ginzburg-Landau parameter, $\kappa$ < 1/$\sqrt{2}$ strongly infers that Pb$_{2}$Pd is a type I superconductor.

\end{abstract}
\maketitle
\section{Introduction}

In recent years, the discovery and study of exotic phases of matter such as Weyl and Dirac semimetal \cite{weyl&dirac}, topological insulators \cite{topIns} and, topological superconductors \cite{topSC} continues to interest both theorists and experimentalists. Among these phases, topological superconductivity (TSC) is one of the most studied phenomena and has become a rapidly developing field in which significant advances have been recently made \cite{topSC}. It not only enriches the existing theoretical framework of physics but also provides a background for investigations on low energy excitations, Majorana fermions which have potential applications in fault-tolerant quantum computation \cite{Majo1,Majo2}. Apart from computation application, TSC itself is an intriguing topological phase of matter, namely, a new type of unconventional superconductivity. In order to realize a topological superconducting state, strong spin-orbital coupling (SOC) is considered one of the necessary ingredients \cite{SOCtop}. Superconducting compounds containing heavy atomic number (Z) elements are proposed to have strong SOC since the strength of SOC is proportional to Z$^{4}$. These superconducting systems with strong SOC often show various exotic properties such as robustness of superconductivity beyond Pauli limit \cite{Pauli1,Pauli2,Pauli3}, nodes and multiple gaps in the superconducting state \cite{nod&mult1,nod&mult2,nod&mult3}, time-reversal symmetry breaking (TRSB) \cite{TRSB1,TRSB2,TRSB3}, as well as being a platform to realize topological superconductivity. Intrinsic topological superconductivity is still a rare phenomenon though various methods have been employed in its study.

To date, Pb rich compounds have not been investigated as potential candidates to exhibit the exotic superconducting features although they display a range of other interesting electronic properties. Recent work on some of the Pb rich compounds (MPb$_{2}$ where M = Au, Rh, Er) predicted them to be a Dirac semimetal, topological superconductor, and coexistence of superconductivity with antiferromagnetic order, respectively \cite{SOCtop3&AuPb2,AuPb2,RhPb2,ErPb2}. The existence of topological phases of matter in Pb rich compounds inspires us to study the superconducting and normal state properties of a theoretically proposed topological material Pb$_{2}$Pd \cite{Pb2Pd_theor,Pb2Pd}.

In this paper, we report the superconductivity in a single crystal of a new type I superconductor Pb$_{2}$Pd. This behaviour is surprising because only a few binary and ternary compounds are known to exhibit type I superconductivity. Electrical transport, magnetization and specific heat measurements confirm bulk superconductivity at a transition temperature, T$_{C}$ = 3.0 $\pm$ 0.1 K. Magnetization and TF-$\mu$SR measurements reveal that Pb$_{2}$Pd is a type~I superconductor. Specific heat measurement indicates a moderate electron-phonon coupling.

\section{Experimental Details}
A Polycrystalline sample of Pb$_{2}$Pd was first synthesized by adding starting materials, Pb and Pd in a ratio of 2:1 in an evacuated sealed quartz tube. It was then heated beyond the melting point (462$\degree$C). A single crystal was prepared by the Bridgeman technique in which the polycrystalline sample was placed in a conical quartz ampoule, heated to 570$\degree$C, dwelled for 30 hours and then slowly cooled to 456$\degree$C, followed by water quenching. Phase purity and crystal structure characterization were performed using a CuK$_{\alpha}$ (1.5406 \AA) equipped PANalytical powder X-ray diffractometer. The quality control and orientation of single crystals were determined using a white beam X-ray diffraction Laue instrument. The data was collected using North Star software. Magnetization and AC susceptibility measurements were carried out in a range of 1.8-10 K and 0-30 mT using a Quantum Design magnetic property measurement system (MPMS 3). Electrical resistivity using a four-probe ac technique in the range 1.9-300 K and specific heat (two-tau relaxation method) measurements were performed on a Quantum Design Physical Property Measurement System (PPMS). 

Transverse field muon spin rotation (TF-$\mu$SR) measurements  were carried out to investigate the magnetic field distribution inside the sample. The data were collected using the M15 beamline at TRIUMF, Centre for Molecular and Material Science, Vancouver, Canada. The spectrometer incorporates a dilution refrigerator which allows for measurements in the low-temperature range till 0.025 K and has a time resolution of 0.4 ns. The single crystal sample used for our measurement was cut into plates before mounting on the cold finger so that a large fraction of muon beam is covered. In TF geometry, the magnetic field was applied parallel to the direction of muon beam whereas measurements were taken with the initial muon spin arranged perpendicular to the field. The direction of the applied magnetic field was perpendicular to the c-axis of crystal. The $\mu$SRfit software package was used to analyze the $\mu$SR data.

\section{Results and Discussion}
\subsection{Crystal Structure characterization}

\begin{figure} [t]
\includegraphics[width=1.0\columnwidth, origin=b]{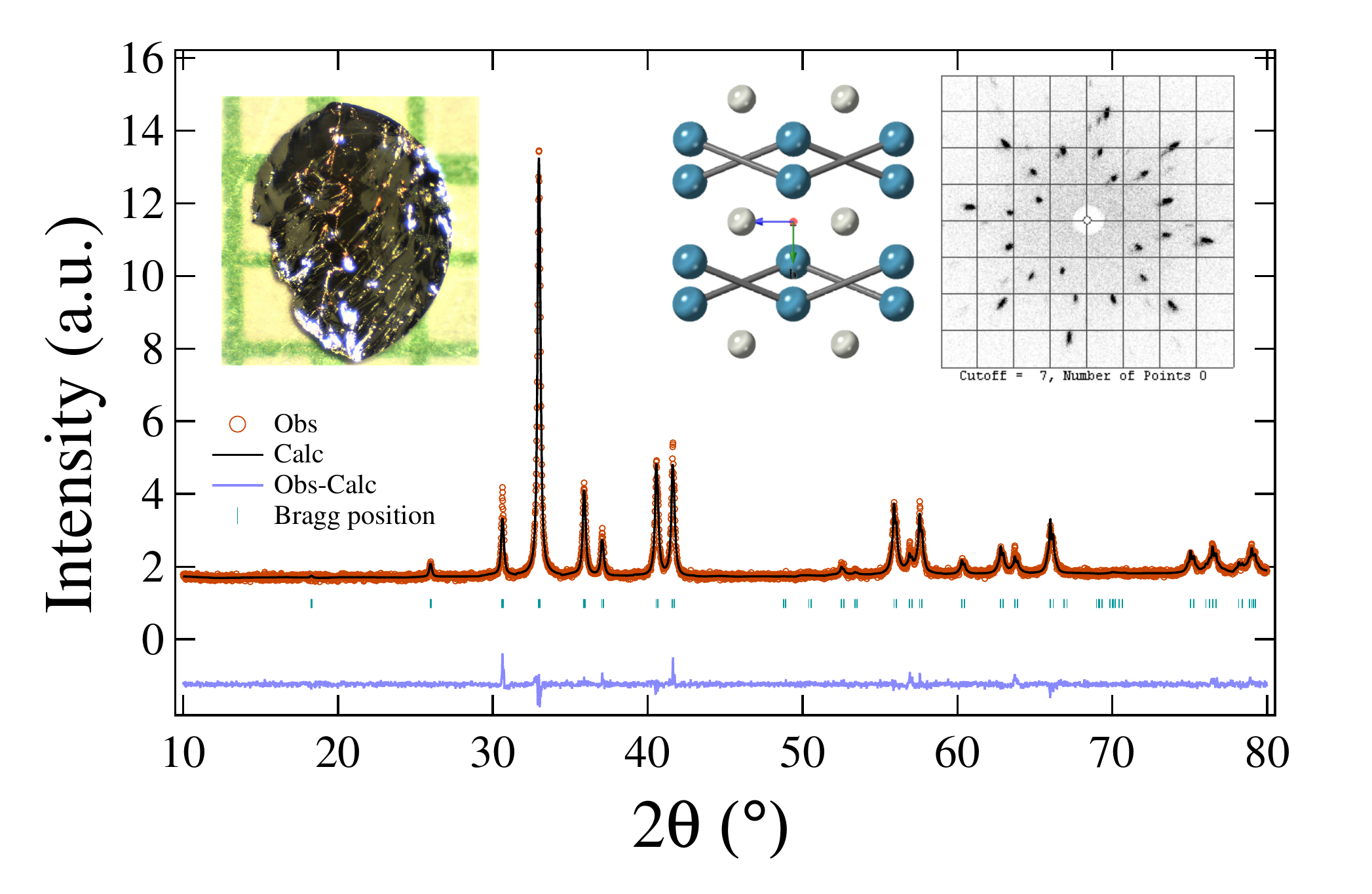}
\caption{\label{Fig1:XRD} Powder XRD pattern of Pb$_{2}$Pd obtained at room temperature is shown by red circles whereas black line represents the Rietveld refinement. Crystal image (left-up) and crystal structure of Pb$_{2}$Pd where blue circles represent Pb atoms, whereas grey circles represent Pd atom (right-up). Back Laue X-ray diffraction pattern of Pb$_{2}$Pd single crystal (right most).}
\end{figure}

Room-temperature X-ray diffraction (XRD) pattern on powdered crystals is shown in \figref{Fig1:XRD}. Rietveld refinement confirms the phase purity, and crystal structure as a body-centred tetragonal with space group I4/mcm. The obtained lattice parameters are as follows: a = b = 6.859 \text{\AA} and c = 5.838 \text{\AA} \cite{P2P_PD1}. All parameters including atomic and Wyckoff positions, cell volume etc. obtained from refinement are summarized in the Table I. \figref{Fig1:XRD}(right) also shows the Laue pattern where the symmetric and bright spots indicate the single-crystalline nature of Pb$_{2}$Pd and correspond to (001) growth direction.

\begin{table}[h!]
\caption{Parameters obtained from Rietveld refinement}
\begin{tabular}{c r} \hline\hline
Structure& Tetragonal\\
Space group&        I4/$mcm$\\ [1ex]
Lattice parameters\\ \hline 
a = b (\text{\AA})&  6.859(3)\\
c (\text{\AA})&  5.838(3)\\

\end{tabular}
\\[1ex]
\begingroup
\setlength{\tabcolsep}{4pt}
\begin{tabular}[b]{c c c c c c}
Atom&  Wyckoff position& x& y& z& \\[1ex]
\hline
Pd1& 4a& 0& 0& 0.250& \\ 
Pb1& 8h& 0.165& 0.665& 0& \\
[1ex]
\hline
\end{tabular}
\par\medskip\footnotesize
\endgroup
\end{table}
\begin{figure} 
\includegraphics[ width=1.0\columnwidth, origin=b]{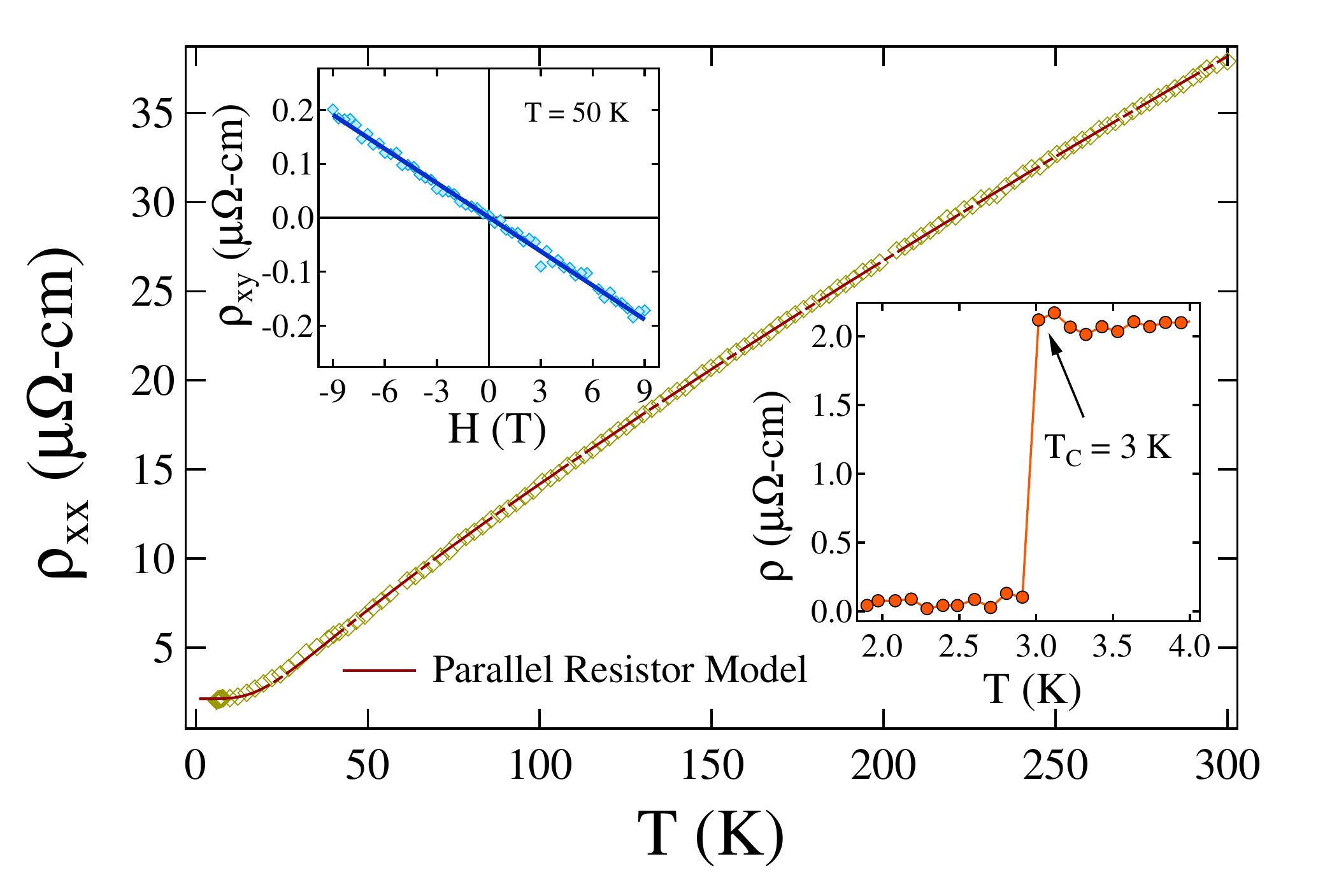}
\caption{\label{Fig2:Rho} Zero-field temperature variation of electrical resistivity in the range 1.9 K $\leq T \leq$ 300 K where solid red line is a fit to parallel resistor model. Inset (bottom right): A sharp superconducting transition at T$_{C}$ = 3.0 K whereas top left inset shows the hall resistivity w.r.t. field ($\pm$ 9T) at T = 50 K.}
\label{fig2}
\end{figure}

\begin{figure*}[t]
\includegraphics[width=2.0\columnwidth,origin=b]{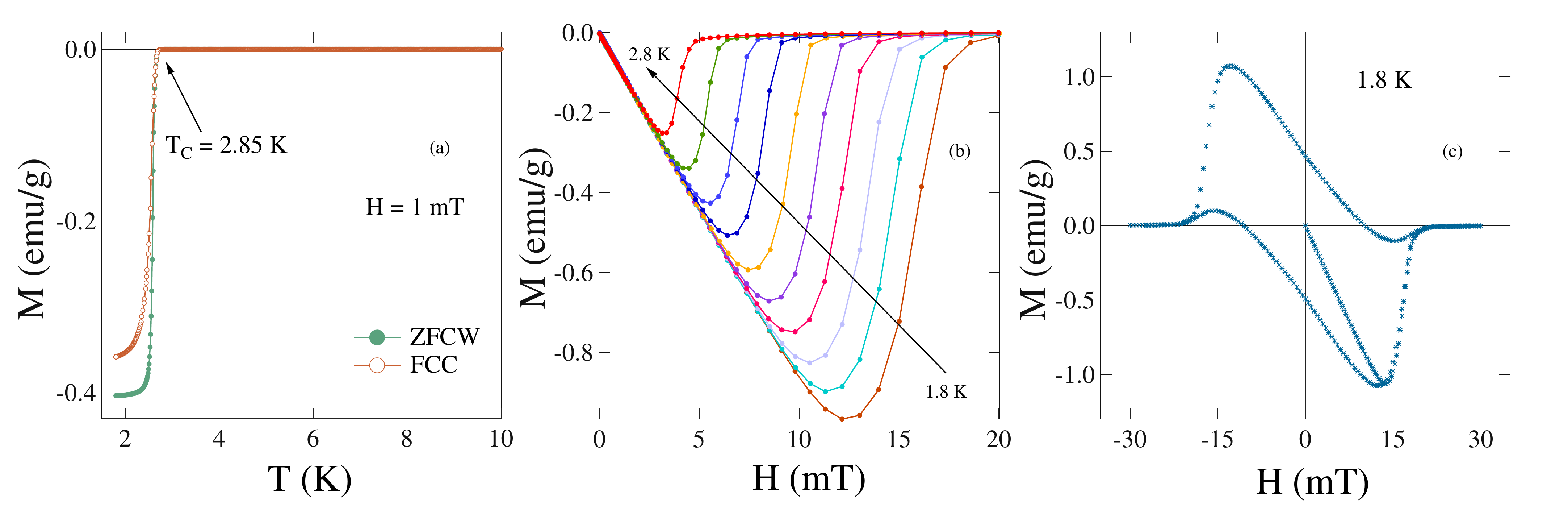}
\caption{\label{Fig3:Mag} a) Temperature dependence of moment collected via ZFCW and FCC under an applied magnetic field of 1 mT. b) Magnetization as a function of applied field for 1.8 K $\leq$ T $\leq$ 2.8 K. c) Magnetization vs magnetic field at 1.8 K of Pb$_{2}$Pd.}
\end{figure*}

\subsection{Normal and superconducting state properties}

\figref{Fig2:Rho} shows the electrical resistivity of Pb$_{2}$Pd as a function of temperature obtained at zero field in the range 1.9-300 K. Inset shows a drop in resistivity at a temperature T$_{c,onset}$ = 3.0 $\pm$ 0.1 K with a transition width $\Delta$T = 0.1 K which confirms superconductivity in this compound. The residual resistivity ratio, (RRR = $\rho(300 K)$/$\rho(6 K)$) = 19 and the low value of $\rho_{0}$ imply a low amount of disorder in the present sample. The saturating type behaviour of $\rho(T)$ instead of linear T dependence at high temperatures, allows this data to be fit by the parallel resistor model. This possibility arises when the order of mean free path becomes short, equal to a few inter-atomic spacing \cite{InterSpac}. In such a case, the scattering cross-section will no longer be linear in scattering perturbation. This leads to the dominant temperature-dependent scattering mechanism is of electron-phonon type interaction at high temperatures. Hence the resistivity will rise less rapidly with temperature showing a saturating behaviour. This type of behaviour in $\rho(T)$ was described by Wiesmann et al. \cite{Parallel} by the following expression
\begin{equation}
 \rho(T) = \left[\frac{1}{\rho_{s}} + \frac{1}{\rho_{i}(T)} \right]^{-1}
\label{para1}
\end{equation}
where $\rho_{s}$ is the temperature-independent saturation resistivity attained at higher temperatures and $\rho_{i}(T)$ is given by
\begin{equation}
 \rho_{i}(T) = \rho_{i,0} + C\left(\frac{T}{\Theta_{D}}\right)^{5}\int_{0}^{\Theta_{D}/T}\frac{x^{5}}{(e^{x}-1)(1-e^{-x})}dx
\label{para2}
\end{equation}

where $\rho_{i,0}$ represent temperature-independent residual resistivity due to scattering from defects in the crystal structure whereas the second term is temperature-dependent generalized BG resistivity in which C is a material-dependent quantity and $\Theta_{D}$ is the Debye temperature obtained from resistivity measurements. The red dashed line in \figref{Fig2:Rho} shows the best fit to the data and yields $\rho_{0}$ = 2.14 $\pm$ 0.03 $\mu\Omega $ cm, C = 63 $\pm$ 1 $\mu\Omega $ cm, $\rho_{0,s}$ = 378 $\pm$ 12 $\mu\Omega $ cm, and $\Theta_{D}$ = 116 $\pm$ 2 K which is close with the $\Theta_{D}$ obtained from specific heat measurement (discussed later).

In order to extract the information regarding the carrier density of Pb$_{2}$Pd, hall resistivity, $\rho_{xy}$ as a function of magnetic field in the range $\pm$ 9 T and at T = 50 K was measured as shown in the top left panel of \figref{Fig2:Rho}. We have determined normal state hall coefficient, R$_{H}$ form the slope of a linear fit to the $\rho_{xy}$(H) data in both field directions. The value obtained for R$_{H}$ = -(2.11 $\pm$ 0.02)$\times10^{-10} \Omega$ m T$^{-1}$ and the negative sign indicate towards the electron as the charge carriers. By using R$_{H}$ = $-1/ne$, where $n$ is the carrier density, and e is the electronic charge, we obtained $n$ = (2.96 $\pm$ 0.02)$\times10^{28} m^{-3}$. Theoretically, the carrier density is defined as n = Z/V$_{cell}$, where Z = 8 is the number of conduction electrons per unit cell and V$_{cell}$ = 0.275$\times10^{-27} m^{3}$ for Pb$_{2}$Pd, yielding a value of n = 2.90 $\times 10^{28}$ m$^{-3}$ and is close to the experimentally obtained value.

Magnetization measurements has been performed on the single crystal oriented along the (001) direction in both parallel and perpendicular magnetic field configuration.  It was obtained via two modes: zero-field cooled cooling (ZFCW) and field cooled cooling (FCC) under an applied magnetic field of 1.0 mT which confirms the superconducting nature of Pb$_{2}$Pd displayed in \figref{Fig3:Mag}(a). A strong diamagnetic signal on entering the superconducting state at a transition temperature of T$_{c,onset}$ = 2.85 $\pm$ 0.10 K is observed, in agreement with previous report \cite{P2P_Tc}. There was no anisotropy in the observed T$_{C}$ in both configurations so rest of the measurements were performed with magnetic field perpendicular to the (001) direction. The Meissner volume fraction indicates the complete expulsion of flux in the superconducting state of this compound. \figref{Fig3:Mag}(c) represents the M-H curve obtained at 1.8 K where it exhibits a diamagnetic signal on re-entering the superconducting state. This type of magnetic behaviour is expected for type I superconductors and has been observed in other type I superconductors including AuBe \cite{AuBe}, Ir$_{2}$Ga$_{9}$ \cite{IrGa} and LaRhSi$_{3}$ \cite{LaRhSi}. The absence of a sharp transition to the normal state is accounted for the demagnetization factor. Low-field magnetization curves obtained for different temperatures (as shown in \figref{Fig3:Mag}(b)) together with Specific heat, and resistivity data at various fields are used to depict the critical field value at 0 K and fit according to
\\
\begin{equation}
H_{C}(T)=H_{C}(0)\left[1-\left(\frac{T}{T_{C}}\right)^{2}\right] .
\label{Hc1}
\end{equation} 
\\ 
The green solid curve shows the Ginzburg-Landau fit and the obtained value for H$_{C}$(0) = 33.6 $\pm$ 1.2 mT.

\begin{figure} 
\includegraphics[width=1.0\columnwidth, origin=b]{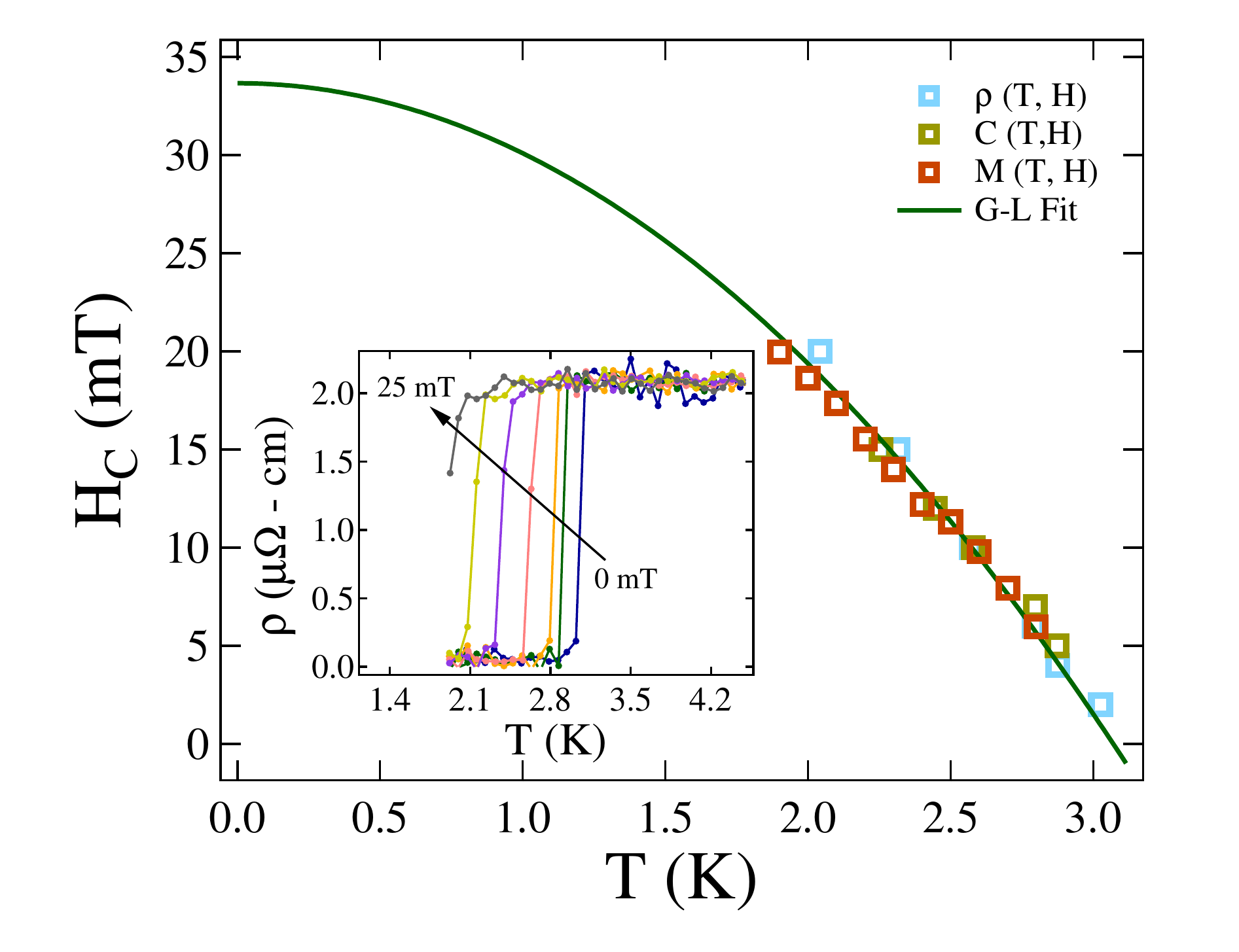}
\caption{\label{Fig4:Hc} Temperature dependence of the critical field from various measurements. Inset: $\rho$(T) at different fields starting from 0 mT to 25 mT.}
\end{figure}

\begin{figure} 
\includegraphics[width=1.0\columnwidth,origin=b]{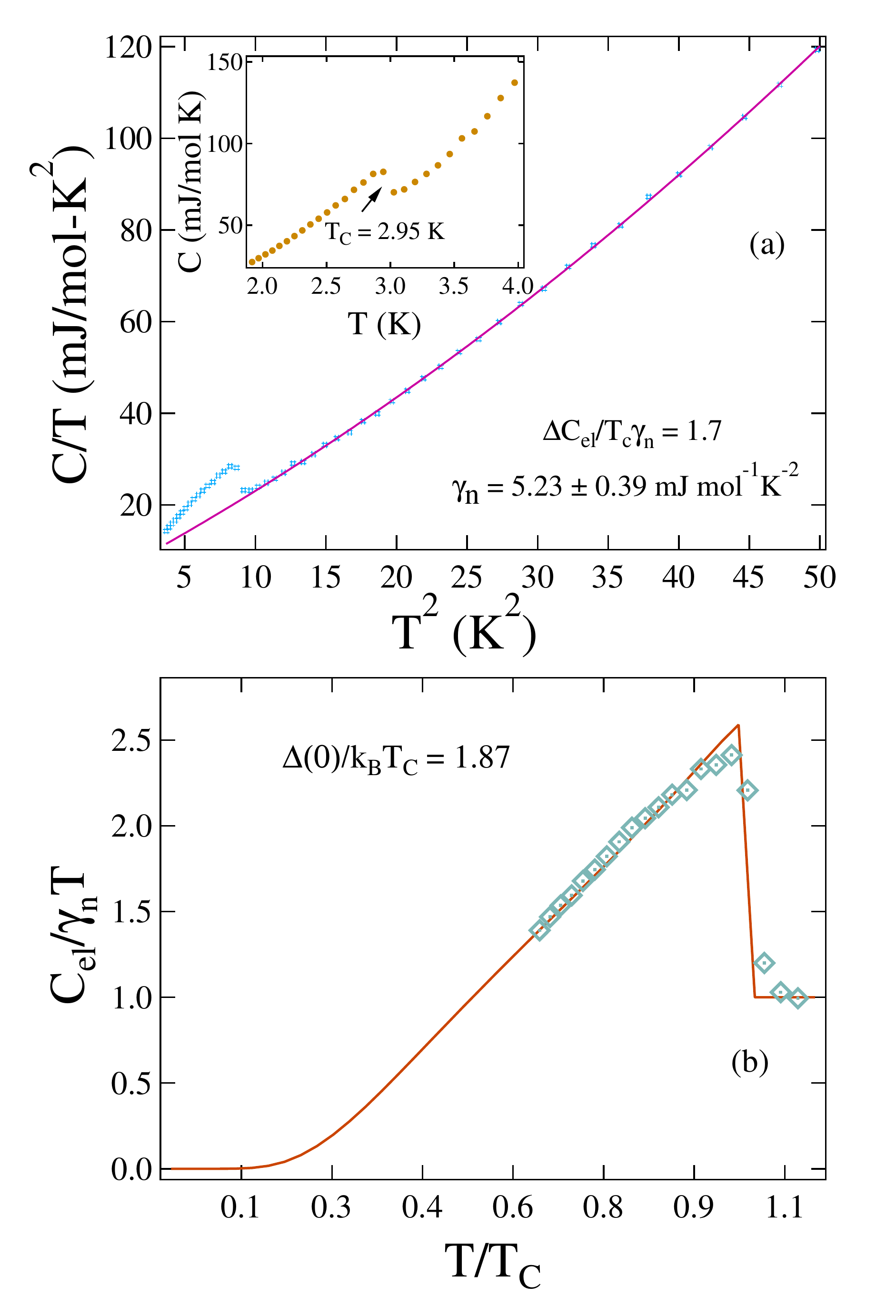}
\caption{\label{Fig5:SH} Zero-field C/T variation with respect to T$^{2}$ where solid line represents a fit to $\frac{C}{T}=\gamma_{n}+\beta_{3} T^{2} + \beta_{5}T^{4}$. Inset: Specific heat data exhibiting a superconducting transition at T$_{C}$ = 2.95 K. b) s-wave fitting to the normalized electronic specific heat data shown by the solid line.}
\end{figure}

Specific heat measurements on the single crystal of Pb$_{2}$Pd is carried out in zero-field from 1.9 K to 10 K and are displayed in the inset of \figref{Fig5:SH}(a). The jump in the specific heat at T$_{c,midpoint}$ = 2.95 $\pm$ 0.10 K which is in agreement with the other measurements confirms the bulk nature of the superconductivity. Above the transition temperature, data is well described by $\frac{C}{T}=\gamma_{n}+\beta_{3} T^{2} + \beta_{5}T^{4}$, shown by the solid pink line in \figref{Fig5:SH}(a) and yields a value of Sommerfeld coefficient $\gamma_{n}$ = 5.23 $\pm$ 0.39 mJ mol$^{-1}$K$^{-2}$, $\beta_{3}$ = 1.65 $\pm$ 0.03 mJ mol$^{-1}$K$^{-4}$, and $\beta_{5}$ = 0.0129 $\pm$ 0.0005 mJ mol$^{-1}$K$^{-6}$. The Debye temperature $\theta_{D}$ \cite{thetaD} is given by $({12\pi^{4}RN}/{5\beta_{3}})^{1/3}$ where using R = 8.314 J mol$^{-1}$K$^{-2}$, N = 3, and $\beta_{3}$ = 1.76 $\pm$ 0.10 mJ mol$^{-1}$K$^{-4}$, we have determined $\theta_{D}$ = 152 $\pm$ 1 K. Further, the Sommerfeld coefficient is related to the density of states at the Fermi energy by $(\pi^{2}k_{B}^{2}D_{C}(E_{F})/{3})$ which yields D$_{C}(E_{F})$ = 2.2 $\pm$ 0.1 states eV$^{-1}$f.u.$^{-1}$. The value of $\theta_{D}$ = 152 $\pm$ 1 K and T$_{c,midpoint}$ = 2.95 $\pm$ 0.10 K are used to estimate $\lambda_{e-ph}$ using McMillian's relation given below \cite{McMillian} 
\\
\begin{equation}
\lambda_{e-ph} = \frac{1.04+\mu^{*}\mathrm{ln}(\theta_{D}/1.45T_{C})}{(1-0.62\mu^{*})\mathrm{ln}(\theta_{D}/1.45T_{C})-1.04}
\label{eqn8:ld}
\end{equation}
\\
where $\mu^{*}$ represents screened repulsive Coulomb potential and usually taken to be 0.13, yielding $\lambda_{e-ph}$ = 0.67 $\pm$ 0.08  which is slightly higher than the other weakly coupled superconductors.

The electronic specific heat (C$_{el}$) contribution has been calculated by subtracting the lattice and phononic contribution from the total specific heat (C). The magnitude of specific heat jump $\Delta C_{el}$/$\gamma_{n}T_{C}$ = 1.7, which is higher than the BCS weak coupling limit 1.43. The value of the same together with $\lambda_{e-ph}$ suggests a moderately coupled superconductivity in Pb$_{2}$Pd. The BCS expression \cite{SH_BCS} for the specific heat below and above T$_{C}$ is described well with the normalized electronic specific heat data and gives a value of $\alpha$ = $\Delta(0)/k_{B}T_{C}$ = 1.87 which is also a bit higher than the BCS value, $\alpha$ = 1.76. Thus, the electronic specific heat data provide compelling evidence of a moderately coupled isotropic s-wave BCS superconductivity in Pb$_{2}$Pd.

We now evaluate various superconducting parameters: $l$ (mean free path), $\lambda_{L}$, $\xi_{0}$, and $\kappa$. Residual resistivity value along with electron carrier density is used to estimate the electronic mean free path, $l$ = v$_{F}\tau$ \cite{m*&Drude} where v$_{F}$ and $\tau$ are the Fermi velocity and the scattering time. Based on the Drude model, the Fermi velocity is given by $\hbar k_{F}$/m$^{*}$ and scattering time $\tau^{-1}$ = ne$^{2}\rho_{0}$/m$^{*}$ where m$^{*}$ is the effective mass and n is the carrier density. The Fermi wave vector is calculated by assuming a spherical Fermi surface, using the following relation k$_{F}$ = (3$\pi^{2}$n)$^{1/3}$, where $n$ = (2.96 $\pm$ 0.02)$\times10^{28} m^{-3}$ was used and k$_{F}$ is estimated to be 0.957~\text{\AA}$^{-1}$. The Sommerfeld coefficient from the specific heat ($\gamma_{n}$ = 5.33 mJmol$^{-1}$K$^{-2}$) and carrier density are used to estimate the effective mass m$^{*}$ = ($\hbar k_{F})^{2}\gamma_{n}$/$\pi^{2}nk_{B}^{2}$ \cite{m*&Drude}, and it provides a value of 2.59 m$_{e}$. By using all the calculated values of n, m$^{*}$, $\rho_{0}$ and k$_{F}$, we determined v$_{F}$ = 4.29$\times10^{5}$ ms$^{-1}$ and $l$ = 638 \text{\AA}.

In order to confirm type I or type II superconductivity, we further investigate London's penetration depth ($\lambda_{L}$), BCS coherence length ($\xi_{0}$) and Ginzburg-Landau parameter ($\kappa_{GL}$). The penetration depth $\lambda_{L}$ = $({m^{*}}/{\mu_{0}n e^{2}})^{1/2}$ is estimated to be 497 \text{\AA} by using the value of m$^{*}$ = 2.59 m$_{e}$ and n = 2.96$\times10^{28}$ m$^{-3}$. Within BCS theory, the coherence length ($\xi_{0}$) is given by  0.18$\hbar v_{F}$/k$_{B}$T$_{C}$, yielding a value of 2008 \text{\AA}. The ratio $l$/$\xi_{0}$ = 0.32 concludes moderately dirty limit superconductivity in Pb$_{2}$Pd. The value of $\kappa_{GL}$ distinguishes between a type I and type II superconductor. In dirty limit, $\kappa_{GL}$ is given by 0.715$\lambda_{L}$(0)/$l$ \cite{kGL} which is found out to be 0.55, less than 1/$\sqrt{2}$, classifies Pb$_{2}$Pd as a type I superconductor. $\kappa_{GL}$ is further determined using the relation 7.49$\times10^{3} \rho_{0}\sqrt\gamma_{nV}$ and found out to be 0.56 by considering $\gamma_{nV}$, $\rho_{0}$ in cgs units, consistent with the value obtained before. To determine Ginzburg-Landau coherence length $\xi(0)$ which is defined as $\lambda_{eff}$/$\kappa_{GL}$ where $\lambda_{eff}$ = $\lambda_{L}({1+\xi_{0}/l})^{1/2}$ = 1011.2 \text{\AA}. Considering $\gamma_{nV}$, $\rho_{0}$ and $\kappa_{GL}$ = 0.55, $\xi(0)$ is evaluated to be 1817 \text{\AA}. 

\begin{figure}
\includegraphics[width=1.0\columnwidth]{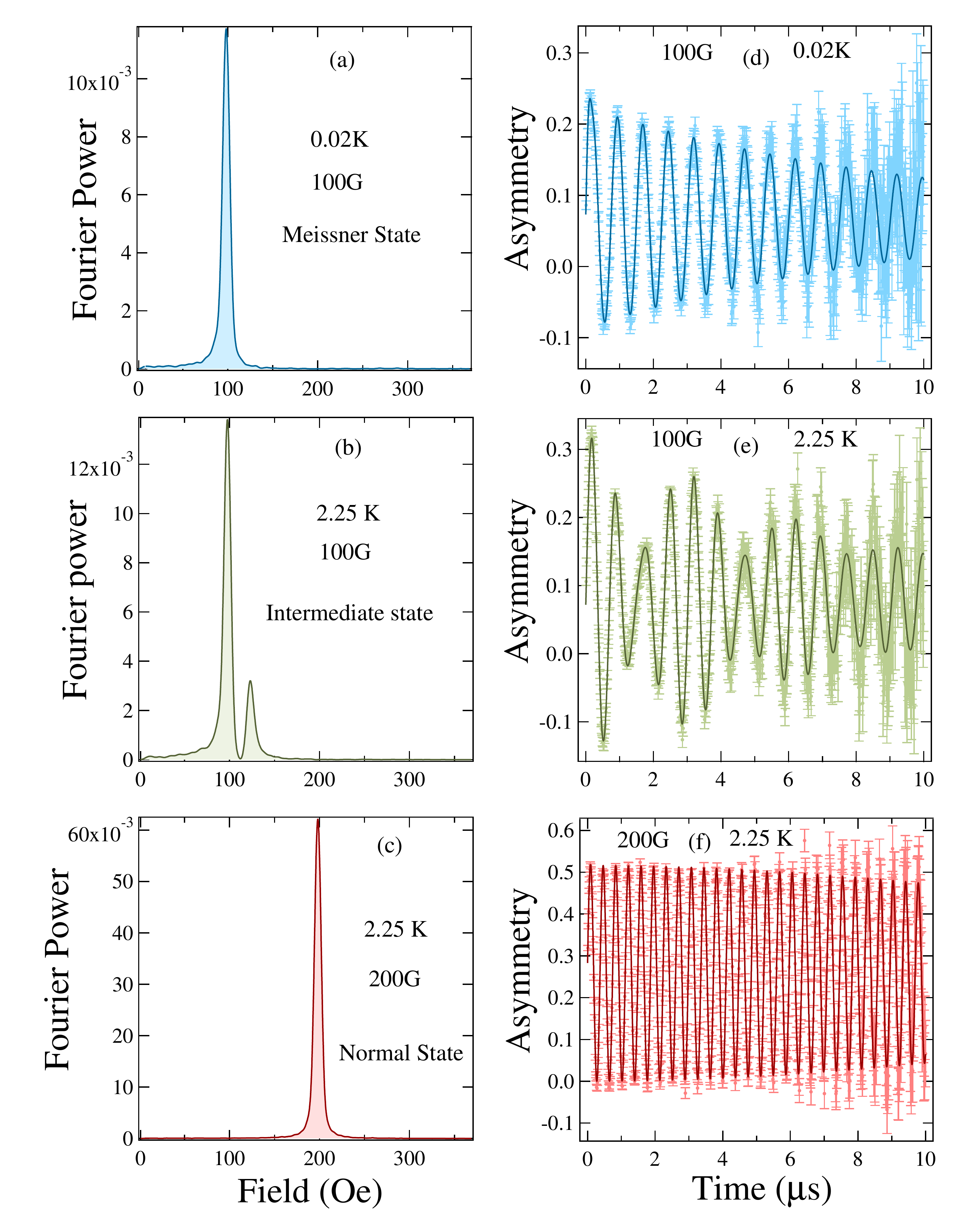}
\caption{\label{Fig6:tf1} The Fourier transform of $\mu$SR asymmetry representing the local field distribution and asymmetry spectra at different fields, and temperatures. (a)-(c) shows field distribution in the Meissner, intermediate and normal state while (d)-(f) consists of corresponding asymmetry spectra. The solid lines are fit to the data using \equref{eqn1:Tranf}.}
\end{figure}

\begin{figure}
\includegraphics[width=1.0\columnwidth]{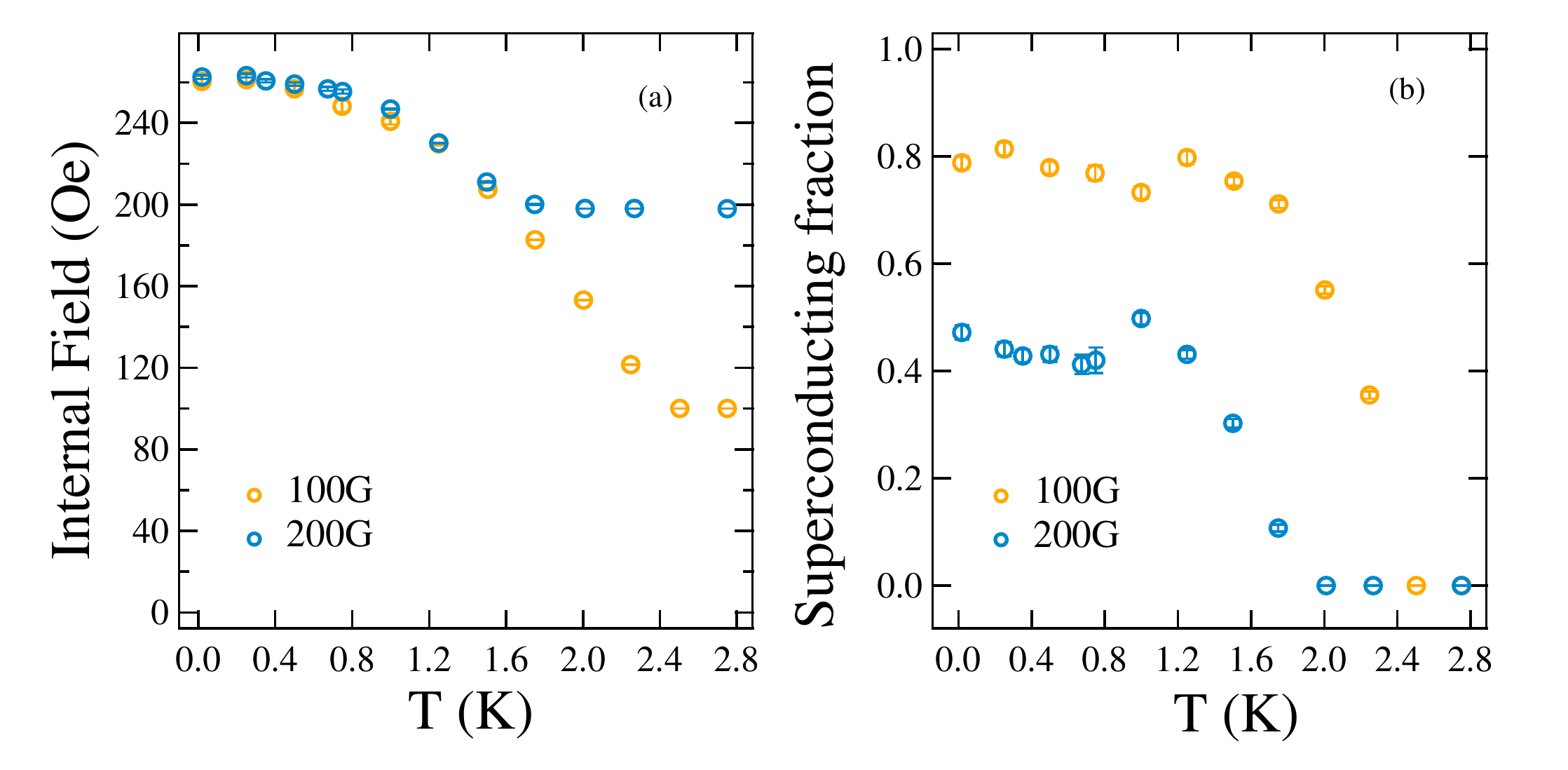}
\caption{\label{Fig7:tf2} a) Internal field as a function of temperature in the normal region of Pb$_{2}$Pd. b) Superconducting volume fraction with respect to temperature at two different fields.} 
\end{figure}

Figure \ref{Fig6:tf1} shows the TF-$\mu$SR spectra and the corresponding internal field distribution in i) Meissner state, ii) Intermediate state, and iii) normal state. The data was collected after field cooling the sample from normal to superconducting state in two different applied fields. The Fourier transform extracted from TF-$\mu$SR spectra provides the information regarding the probability distribution of an internal field in the intermediate state of a type-I or a vortex state of a type-II superconductor. The presence of only one peak at H = H$_{app}$ = 100 G and T = 0.02 K corresponds to complete superconducting or Meissner state where all the magnetic flux gets expelled from the inside of a superconductor. The peak at H = 100 G represents the background signal originating due to muons stopping in the sample holder. At T = 2.25 K and H$_{app}$ = 100 G shows the intermediate state of Pb$_{2}$Pd. The asymmetry spectra were analyzed using the following equation:
\begin{equation}
\begin{split}
A (t) = A_{0}[F(1-F_{S})\cos(\gamma_\mu H_N+\phi)\exp\left(-\frac{1}{2}\sigma_N^2t^2\right)\\ + (1-F)\cos(\gamma_\mu H_{bg}+\phi)\exp\left(-\lambda_{bg} t\right)],
\label{eqn1:Tranf}
\end{split}
\end{equation}
where A$_{0}$ is the initial asymmetry, F$_{S}$ and F are the superconducting and total fraction coming from the sample, H$_{N}$ is the internal field in the normal regions of the sample, H$_{bg}$ is the contribution from the background field which originates due to muon stopping in the sample holder, $\phi$ is the phase shift, $\sigma_{N}$ is the normal region relaxation rate, and $\lambda_{bg}$ is the background relaxation rate. The fit using \equref{eqn1:Tranf} at T = 2.25 K and H = 100 G provide two values of field: 100 G and 123 G. The first value corresponds to the applied field while the second one is taken as an estimate of the critical field in the intermediate state of a type I superconductor which is induced by the non-zero demagnetization effects. This causes some part of the superconductor to experience a field greater than the applied field and gives rise to the coexistence of superconducting and normal regions even if the applied magnetic field is less than the critical field. Muons landing in the superconducting region will not precess whereas muons implanted in the normal region will precess with a magnetic field $\sim$ H$_{C}$. The Fourier transform shown in \figref{Fig6:tf1} (b) consists of two peaks in which one peak at H = 100 G corresponds to the muon stopping in the silver cold finger as well as the normal state regions of the superconductor whereas the short peak centered at H$_{int}$ = 123 G corresponds to the intermediate state in Pb$_{2}$Pd. This unambiguously demonstrates that Pb$_{2}$Pd is a type-I superconductor as the mixed state in type-II superconductor will have a lower field value than the applied field due to the establishment of flux line lattice. 

The $\mu$SR asymmetry spectrum was also analyzed as a function of temperature at 100 and 200 G fields. Figure \ref{Fig7:tf2} (a) shows the temperature dependence of the internal field in the normal region of Pb$_{2}$Pd.  At T << T$_{C}$ and H$_{app}$ < H$_{C}$, the internal field is equal to the critical field and as the temperature increases it approaches the H$_{app}$ value. Superconducting volume fraction with respect to temperature is shown in \figref{Fig7:tf2} (b) and it shows an increasing behaviour as the temperature decreases as it is expected for a type I superconductor. 
\begin{figure}
\includegraphics[width=1.0\columnwidth]{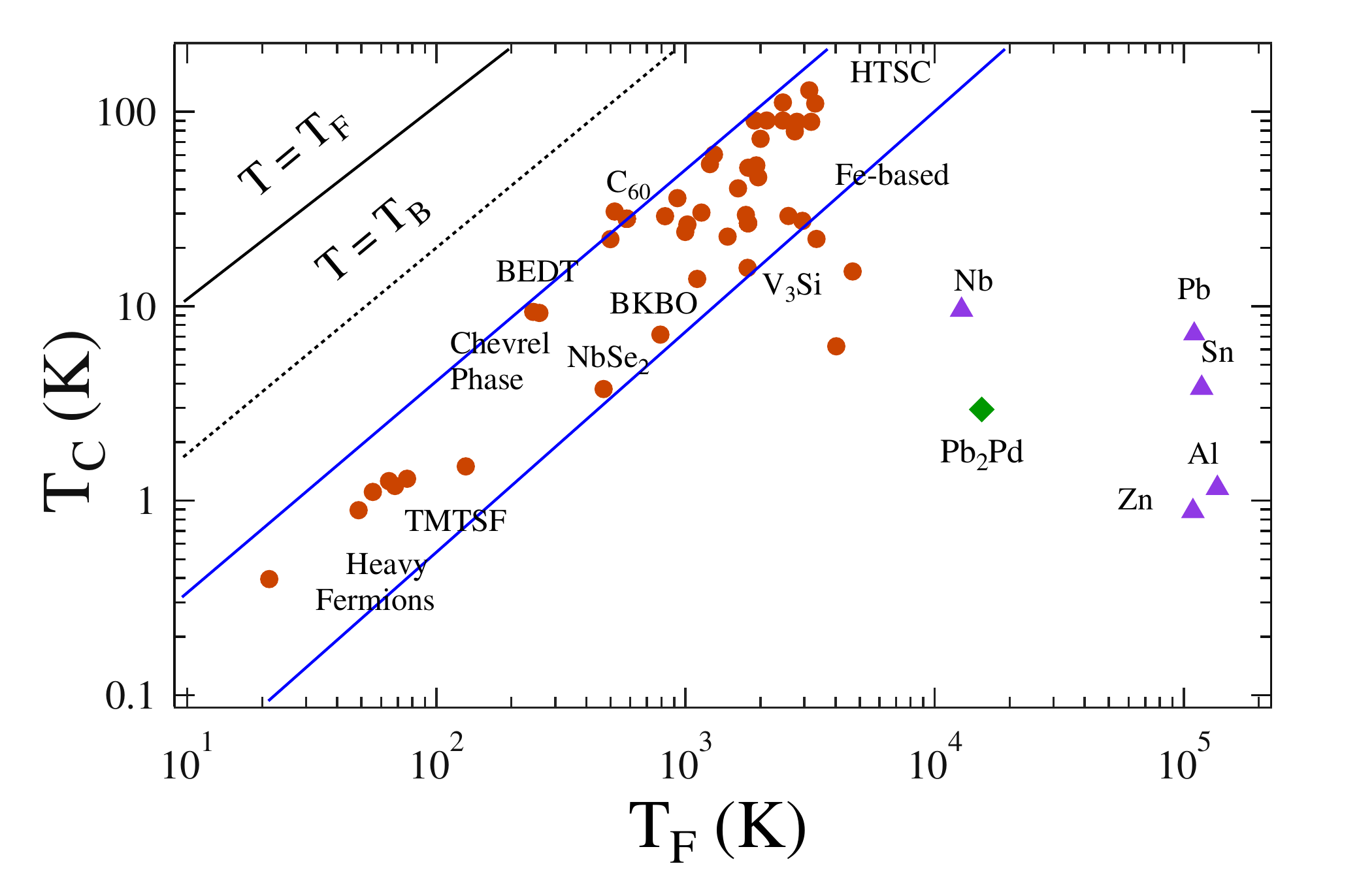}
\caption{\label{Fig8:Uemura} The Uemura plot showing the superconducting transition temperature $T_{c}$ vs the effective Fermi temperature $T_{F}$, where Pb$_{2}$Pd is shown as a solid red marker. Other data points plotted between the blue solid lines is the different families of unconventional superconductors \cite{Umera_1,Umera_2,Umera_3}.} 
\end{figure}

\begin{table}
\caption{Superconducting and normal state parameters of Pb$_{2}$Pd}

\begin{tabular}[b]{l c c }\hline\hline
Parameters& unit& Pb$_{2}$Pd\\
\hline
$T_{C}$& K& 3.0 (1)\\             
$H_{C}(0)$& mT& 34 (1)\\
$\Delta C_{el}/\gamma_{n}T_{C}$&   &1.7 (1)\\
$\theta_{D}$& K& 152 (1)\\
$\lambda_{e-ph}$& & 0.7 (1)\\
$\Delta(0)/k_{B}T_{C}$&  &1.9 (2)\\
$\xi(0)$&  \text{\AA}& 1817 (280)\\
$m^{*}/m_{e}$& & 2.6 (1)\\             
n& 10$^{28}$m$^{-3}$& 2.96 (2)\\
$l$&  \text{\AA}& 638 (60)\\ 
$\xi_{0}$&  \text{\AA}& 2008 (220)\\
$\lambda_{L}$& \text{\AA}& 497 (20)\\
$l/\xi_{0}$& & 0.32\\
$k_{GL}$& &0.55 (2)\\
$v_{f}$& 10$^{5}$ms$^{-1}$& 4.3 (1)\\
$T_{F}$& K& 13629\\
$T_{C}/T_{F}$& & 0.00019\\
\hline\hline
\end{tabular}
\par\medskip\footnotesize

\end{table}

To classify a superconductor as a conventional or unconventional, Uemura et al. \cite{Umera_1,Umera_2,Umera_3} provides a classification based on their $\frac{T_{C}}{T_{F}}$ ratio. This ratio falls in the range 0.01 $\leq$ $\frac{T_{C}}{T_{F}}$ $\leq$ 0.1 for many unconventional such as heavy-fermion, iron-based superconductors and high T$_{C}$ cuprates shown between the two blue solid lines. To calculate T$_{F}$ for a 3D system, the following relation is used: $k_{B}T_{F}$ = ($\hbar k_{F})^{2}/{2m^{*}}$ where k$_{F}$ is the Fermi wave vector. This gives T$_{F}$ = 13629 K and the ratio $\frac{T_{C}}{T_{F}}$ = 0.00019. It places Pb$_{2}$Pd away from the band of unconventional superconductors (as shown by a green symbol). But at the same time, Pb$_{2}$Pd is also away from the type I conventional superconductors which are mostly pure elemental superconductors.

\section{CONCLUSION}
In this work, we have examined the superconducting properties of a new superconductor Pb$_{2}$Pd using specific heat, electric transport, magnetization, and TF $\mu$SR measurements and found that it undergoes a sharp superconducting transition at T$_{C}$ = 2.95 K. The value of residual resistivity, full Meissner fraction and small width of superconducting transition in all measurements indicate towards the high sample purity and crystallographic order. The field dependence of magnetization together with other superconducting parameters listed in Table II obtained from Sommerfeld coefficient and the residual resistivity value within the framework of BCS theory strongly revealed a type I  and moderately dirty limit superconductivity in Pb$_{2}$Pd. Low-temperature specific heat measurements suggest isotropic s-wave superconductivity with a moderate electron-phonon coupling. The microscopic study of Pb$_{2}$Pd using TF $\mu$SR measurement confirms type I superconductivity. To establish a relation between the topological nature of the material and superconducting ground-state properties, further microscopic measurements need to be done.   

\section{Acknowledgments} R.~P.~S.\ acknowledges Science and Engineering Research Board, Government of India for the Core Research Grant CRG/2019/001028.


\begin{thebibliography}{References}

 \bibitem{weyl&dirac} N. P. Armitage, E. J. Mele, and A. Vishwanath, Rev. Mod. Phys. 90, 015001 (2018).

 \bibitem{topIns} M. Z. Hasan and C. L. Kane, Rev. of Mod. Phys. 82, 3045 (2010).
 
 \bibitem{topSC} Y. Li and Z.-A. Xu, Adv. Quantum Technol. 2, 1800112 (2019).
 
 \bibitem{Majo1} Jason Alicea, Yuval Oreg, Gil Refael, Felix von Oppen, Matthew P. A. Fisher, Nature Physics 7, 412-417 (2011).
 
 \bibitem{Majo2} X.-J. Liu, C. L. M. Wong, K. T. Law, Phys. Rev. X 4, 021018 (2014).
 
 \bibitem{SOCtop} M. Sato and Y. Ando, Rep. Prog. Phys. 80, 076501 (2017).
 
 \bibitem{Pauli1} M. Uchida, M. Ide, M. Kawamura, K. S. Takahashi, Y. Kozuka, Y. Tokura, M. Kawasaki, Phys. Rev. B 99, 161111(R) (2019).

 \bibitem{Pauli2} N. Kimura, K. Ito, H. Aoki, S. Uji, T. Terashima, Phys. Rev. Lett. 98, 197001 (2007).
 
 \bibitem{Pauli3} E. M. Carnicom, W. W. Xie, T. Klimczuk, J. J. Lin, K. G\'ornicka, Z. Sobczak, N. P. Ong, and R. J. Cava, Sci. Adv. 4, 7969 (2018).

 \bibitem{nod&mult1} K. Maki, S. Haas, D. Parker, H. Won, K. Izawa and Y. Matsuda, Europhys. Lett. 68, 720-725 (2004).

 \bibitem{nod&mult2} K. Izawa, Y. Nakajima, J. Goryo, Y. Matsuda, S. Osaki, H. Sugawara, H. Sato, P. Thalmeier, K. Maki, Phys. Rev. Lett. 90, 117001 (2003).
 
 \bibitem{nod&mult3} Z. F. Weng, J. L. Zhang, M. Smidman, T. Shang, J. Quintanilla, J. F. Annett, M. Nicklas, G. M. Pang, L. Jiao, W. B. Jiang, Y. Chen, F. Steglich, and H. Q. Yuan, Phys. Rev. Lett. 117, 027001 (2016).
 
 \bibitem{TRSB1} Y. Aoki, A. Tsuchiya, T. Kanayama, S. R. Saha, H. Sugawara, H. Sato, W. Higemoto, A. Koda, K. Ohishi, K. Nishiyama, and R. Kadono, Phys. Rev. Lett. 91, 067003 (2003).
 
 \bibitem{TRSB2} G. M. Luke, Y. Fudamoto, K. M. Kojima, M. I. Larkin, J. Merrin, B. Nachumi, Y. J. Uemura, Y. Maeno, Z. Q. Mao, Y. Mori, H. Nakamura and M. Sigrist, Nature 396, 658-660 (1998).
 
 \bibitem{TRSB3} A.D. Hillier, J. Quintanilla, B. Mazidian, J. F. Annett, and R. Cywinski, Phys. Rev. Lett. 109, 097001 (2012).
 
 \bibitem{SOCtop3&AuPb2} Y. Xing, H. Wang, C.-K. Li, X. Zhang, J. Liu, Y. Zhang, J. Luo, Z. Wang, Y. Wang, L. Ling, M. Tian, S. Jia, J. Feng, X.-J. Liu, J. Wei, and J. Wang, npj Quantum Materials 1, 16005 (2016).
 
 \bibitem{AuPb2} L. M. Schoop, L. S. Xie, R. Chen, Q. D. Gibson, S. H. Lapidus, I.r Kimchi, M. Hirschberger, N. Haldolaarachchige, M. N. Ali, C. A. Belvin, T. Liang, J. B. Neaton, N. P. Ong, A. Vishwanath, and R. J. Cava, Phys. Rev. B 91, 214517 (2015).
 
 \bibitem{RhPb2} J.-F. Zhang, P.-J. Guo, M. Gao, K. Liu, and Z.-Y. Lu, Phys. Rev. B 99, 045110 (2019). 
 
 \bibitem{ErPb2} Y. Hattori, K. Fukamichi, T. Goto, K. Suzuki, J. Phys. Soc. Jpn. 61, pp. 3845-3848 (1992).
 
 \bibitem{Pb2Pd_theor} T. Zhang, Y. Jiang, Z. Song, H. Huang, Y. He, Z. Fang, H. Weng, C. Fang, arXiv:1807.08756v1 (2018).
 
 \bibitem{Pb2Pd} M.F. Gendron, R.E. Jones, J. Phys. Chem. Solids 23 405-406 (1962).
 
  \bibitem{P2P_PD1} E. E. Havinga, H. Damsma, P. Hokkeling, Journal of the Less-Common Metals 27, 169-186 (1972).
  
 \bibitem{InterSpac} Z. Fisk and G. W. Webb, Phys. Rev. Lett. 36, 1084 (1976).
 
 \bibitem{Parallel} H. Wiesmann, M. Gurvitch, H. Lutz, A. K. Ghosh, B. Schwarz, M. Strongin, P. B. Allen, and J. W. Halley, Phys. Rev. Lett. 38, 782 (1977).
  
  \bibitem{P2P_Tc} E. E. Havinga, H. Damsma, P. Hokkeling, Journal of the Less-Common Metals 27, 281-291 (1972).
  
  \bibitem{AuBe} D. Singh, A. D. Hillier, R. P. Singh, Phys. Rev. B 99, 134509 (2019)
  
   \bibitem{IrGa} T. Shibayama, M. Nohara, H. A. Katori, Y. Okamoto, Z. Hiroi, and H. Takagi, J. Phys. Soc. Jpn. 76, 073708 (2007).
  
   \bibitem{LaRhSi} V. K. Anand, A. D. Hillier, D. T. Adroja, Phys. Rev. B 83, 064522 (2011).
    
   \bibitem{thetaD}  D. Singh, J. A. T. Barker, A. Thamizhavel, D. McK. Paul, A. D. Hillier, and R. P. Singh, Phys. Rev. B 96, 180501(R) (2017).
  
  \bibitem{McMillian} W. L. McMillan, Phys. Rev. 167, 331 (1968).
  
  \bibitem{SH_BCS} D. Singh, A. D. Hillier, A. Thamizhave, R. P. Singh, Phys. Rev. B 94, 054515 (2016).
  
  \bibitem{m*&Drude} M. Tinkham, Introduction to Superconductivity (McGraw-Hill, New York, 1996).
  
  \bibitem{kGL} M. Tinkham, Introduction to Superconductivity, 2nd ed. (Dover, Mineola, NY, 1996).
  
   \bibitem{Umera_1} Y. J. Uemura, V. J. Emery, A. R. Moodenbaugh, M. Suenaga, D. C. Johnston, A. J. Jacobson, J. T. Lewandowski, J. H. Brewer, R. F. Kiefl, S. R. Kreitzman, G. M. Luke, T. Riseman, C. E. Stronach, W. J. Kossler, J. R. Kempton, X. H. Yu, D. Opie, H. E. Schone, Phys. Rev. B 38, 909(R) (1988).
  
  \bibitem{Umera_2} Y. J. Uemura, G. M. Luke, B. J. Sternlieb, J. H. Brewer, J. F. Carolan, W. N. Hardy, R. Kadono, J. R. Kempton, R. F. Kiefl, S. R. Kreitzman, P. Mulhern, T. M. Riseman, D. Ll. Williams, B. X. Yang, S. Uchida, H. Takagi, J. Gopalakrishnan, A. W. Sleight, M. A. Subramanian, C. L. Chien, M. Z. Cieplak, Gang Xiao, V. Y. Lee, B. W. Statt, C. E. Stronach, W. J. Kossler, X. H. Yu, Phys. Rev. Lett. 62, 2317 (1989).
  
  \bibitem{Umera_3} Y. J. Uemura, L. P. Le, G. M. Luke, B. J. Sternlieb, W. D. Wu, J. H. Brewer, T. M. Riseman, C. L. Seaman, M. B. Maple, M. Ishikawa, D. G. Hinks, J. D. Jorgensen, G. Saito, H. Yamochi, Phys. Rev. Lett. 66, 2665 (1991).
  
\end{thebibliography}
\end{document}